\documentstyle{mn}
\newcommand{\zbar}{\bar z}

\input{epsf.sty}
\input{psfig.sty}

\def\mpc {h^{-1} {\rm{Mpc}}}
\def\and  {\it {et al.} \rm}
\def\rmd {\rm d}
\def\dk{(\Delta k)}
\def\dz{(\Delta z)}
\def\eql#1{\label{eq:#1}}
\def\ec#1{eq.~(\ref{eq:#1})}
\def\Ec#1{(\ref{eq:#1})}

\def\spose#1{\hbox to 0pt{#1\hss}}
\def\simlt{\mathrel{\spose{\lower 3pt\hbox{$\mathchar"218$}}
     \raise 2.0pt\hbox{$\mathchar"13C$}}}
\def\simgt{\mathrel{\spose{\lower 3pt\hbox{$\mathchar"218$}}
     \raise 2.0pt\hbox{$\mathchar"13E$}}}
\def\be{\begin{equation}}
\def\ee{\end{equation}}
\def\bce{\begin{center}}
\def\ece{\end{center}}
\newcommand{\vs}{\nonumber\\}
\def\bea{\begin{eqnarray}}
\def\eea{\end{eqnarray}}
\def\ben{\begin{enumerate}}
\def\een{\end{enumerate}}

\def\brr{\begin{array}}
\def\err{\end{array}}

\def\nh1{n_{\rm HI}}

\def\p1dk{P_{\rm 1D}(k)}
\def\simlt{\mathrel{\spose{\lower 3pt\hbox{$\mathchar"218$}}
     \raise 2.0pt\hbox{$\mathchar"13C$}}}
\def\simgt{\mathrel{\spose{\lower 3pt\hbox{$\mathchar"218$}}
     \raise 2.0pt\hbox{$\mathchar"13E$}}}

\begin{document}

\title{Non-Linear Effects on the Angular Correlation Function}

\author[R.~Scranton and S. Dodelson]{Ryan Scranton$^{1,2}$
and Scott Dodelson$^{1,2}$ 
\vspace{1mm}\\
$^1$ NASA/Fermilab Astrophysics Center, P.O. Box 500, Batavia, IL 60510 USA \\
$^2$ Department of Astronomy and Astrophysics, University of Chicago, Chicago, IL
60637 USA}

\maketitle 
 
\def\mpc {h^{-1} {\rm Mpc}}
\def\impc {h {\rm Mpc}^{-1}}
\def\and  {{\it {et al.} }}
\def\rmd {{\rm d}}

\begin{abstract}

Extracting the three dimensional 
power spectrum from the 2D distribution
of galaxies has become a standard tool of cosmology. 
This extraction requires some assumptions
about the scaling of the power spectrum with redshift;
all treatments to date assume a simple power law
scaling. In reality, different scales grow at different
rates, due to non-linearities. We show that angular
surveys are sensitive to a weighted average of the power spectrum over a distribution
of redshifts, where the weight function varies with wavenumber.
We compute this weight function and
show that it is fairly sharply peaked at $\bar z$,
which is a function of $k$. As long as the 
extracted power spectrum is understood to be
$P(k,\bar z)$, the error introduced by 
non-linear scaling is quantifiable and small.
We study these effects in the context of the APM
and SDSS photometric surveys. In general the
weight matrix is peaked at larger $z$ and is broader
for deeper surveys, leading to larger (but still
quantifiable) errors
due to non-linear scaling.
The tools introduced here -- in particular the weight function and effective
redshift $\zbar$ -- can also be
profitably applied to plan surveys to study the evolution of the power spectrum.

\end{abstract}

\bigskip

%%\begin{keywords}
%\keywords{large scale structure of the universe --- methods: numerical}
%%large scale structure of the universe --- methods: numerical
%%\end{keywords}

\section{Introduction}

Even as redshift surveys which allow us to obtain three
dimensional maps of the sky advance, photometric surveys
still maintain their usefulness for cosmology. The fundamental advatange 
of the two dimensional surveys is that they can measure the
positions of many more galaxies than can redshift surveys. This
is often enough to offset the loss of radial information, especially
if one is interested only in some simple statistics characterizing
the underlying density field. 

In order to make sense of the angular information, one needs to understand
how structure is sampled along the line of sight. For example, a deep
survey picks up information about structure at much earlier times
and much larger scales than a shallow one. Most of this information
is encoded in the {\it kernel} which is given by Limber's Equation
if the selection function is known. Recapturing the three-dimensional
power spectrum from the angular correlation function then involves inverting
the kernel. This inversion is not completely straightforward, but several
different techniques have been used (Baugh \& Efstathiou 1993,1994;
Gazta\~{n}aga \& Baugh 1998; Dodelson \& Gazta\~{n}aga 1999; and 
Eisenstein \& Zaldarriaga 1999), and they all seem to agree fairly well.

One aspect of Limber's equation which is typically given short shrift 
is the question of how the power spectrum (or its Fourier transform,
the correlation function) evolves with time. Some assumption is needed
in order to generate the kernel; typically it has been assumed that the
power spectrum scales as $(1+z)^{-\beta}$ where $z$ is the redshift and
$\beta=2$ corresponding to linear evolution in a flat universe
is the standard choice. It is important to note that making a choice
is crucial to the success of the inversion process. An assumption about
the time dependence of the power spectrum allows the kernel to be written
as a function of wavenumber $k$ only; undoing the integral over $k$
for many angles is then possible. If no assumption was made, the integral would
be over both redshift and $k$. It would be much more difficult, if not
impossible, to undo
this two-dimensional integration and get out $P(k,z)$.

Since we are forced into an assumption about how the power spectrum evolves with
time, we need to ask how much this assumption affects
the results. Here we examine this question. Section 2 briefly reviews the
standard derivation of Limber's Equation. Section 3 introduces a tool
to analyze the effectiveness of the assumption that $P\propto (1+z)^{-\beta}$.
This is the recent work which allows one to generates a full non-linear
$P(k,z)$ from a given linear power spectrum. Armed with this tool, we then
show two power spectra, both of which give the same angular correlation
function. One has the simple $(1+z)^{-\beta}$ scaling, while the other has
more realistic scaling accounting for non-linearities. Although these
two 3D power spectra give the same angular correlation function, they are much
different today. We illustrate this for an APM-like 
survey (Maddux et al. 1990) .

Section 4 isolates the reason for the difference between the two spectra.
Essentially, any survey is actually a measure of the power spectrum
over a range of redshifts centered at $\zbar$, an easily computable function
of wavenumber. If one insists on interpreting the results as measures
of the power spectrum today, different scalings from $\zbar$ to
$z=0$ lead to different $P(k,z=0)$. However, if one interprets the results 
as a measurement of $P(k,\zbar)$, the scaling scheme one uses is
irrelevant. Another way of saying this is to emphasize that the measurement is
of $P(k,\zbar)$; the weighting function is fairly compact and so is
often insensitive to the behavior of $P$ for $z$ much different than $\zbar$.

Finally section 5 computes the error in the estimate of $P(k,\zbar)$
introduced by assuming linear scaling through $\zbar$. This error is
largest on small scales where non-linear evolution sets in earliest. For 3D wave
numbers $k \la 1$ h Mpc$^{-1}$, the error is 
quite small for APM, and larger, but still less than the statistical errors for
a wider, deeper survey such as the Sloan Digital Sky Survey 
(SDSS)\footnote{{\tt http://www.sdss.org}} .

\section{The Kernel and the Standard Assumption}

We begin with the discretized, relativistic version of Limber's equation,
\begin{equation}
w_i = \sum_{a\alpha} {\cal K}_{ia\alpha} P_{a\alpha}
\eql{fullim}\ee
where $w_i$ is the angular correlation function in a bin centered at $\theta_i$;
$P_{a\alpha}$ is the power spectrum in a bin centered at wavenumber $k_a$ and 
redshift $z_\alpha$; and ${\cal K}$ is the kernel which depends on all three
variables. The kernel is
\bea
{\cal K}_{ia\alpha} &=& 
\frac{\dk\dz}{2\pi} {k_a 
{x(z_\alpha)^{4}} \over  (1+z_\alpha)^{6} F(z_\alpha) H(z_\alpha)} 
\vs
&&\times \psi^{2}(z_\alpha) 
J_{0}(k_a x(z_\alpha) \sin \theta_i),
\eql{fullker}\eea
where $x$ is the comoving distance out to redshift $z$; $F$ depends
on the cosmological model, equal to one in a flat, matter dominated universe;
$H$ is the Hubble rate as a function of redshift which is also model dependent;
and $\psi$ is the selection function.
It has been normalized so
 that
\begin{equation}
\int^{\infty}_{0} dx x^{2} a^{3}(x) \psi(x) = 1.
\end{equation}

If one assumes a linearly evolving power spectrum, then 
\begin{equation}
P(k,z) = P(k,z=0) (1+z)^{-2},
\eql{STDGRO}
\end{equation}
and $P(k,0)$ can be moved out of the sum
over redshifts.  In this case, the kernel simplifies and we are left with:
\begin{equation}
w_i = K_{ia} P_{a}
\eql{STDLIM}
\end{equation}
where $P_a$ now denotes the power spectrum today and
the new kernel is independent of redshift:
\bea
K_{ia} &=& \frac{\dk\dz}{2\pi} k_a \sum_\alpha
{x(z_\alpha)^{4} \over (1+z_\alpha)^{8} F(z_\alpha) H(z_\alpha)} 
\vs
&&\times \psi^{2}(z_\alpha) 
J_{0}(k_a x(z_\alpha) \sin \theta_i).
\eea
Equation~\Ec{STDLIM} is then inverted to extract the three dimensional
power spectrum $P_a$.  

To reiterate, the separation in \ec{STDGRO} is certainly not correct, 
since the power on small scales 
evolves differently over time than the 
power at large scales.  We now turn to more realistic scaling.

\section{Non-Linear Scaling}

To arrive at a non-linear power spectrum from a linear one, we use the 
treatment described by Peacock \& Dodds (1996).  In their paper they work in 
terms of the dimensionless power spectrum $\Delta^{2}$, where
\begin{equation}
\Delta^{2}(k) = \frac{k^{3}}{2\pi^{2}} P(k).
\end{equation}
They introduce the non-linear wave-number, $k_{NL}$,
a function of the linear wave-number,
$k_{L}$ and the non-linear power spectrum $\Delta^{2}_{NL}$. In particular,
\begin{equation}
k_{NL} = (1 + \Delta^{2}_{NL}(k_{NL}))^{3} k_{L}
\end{equation}
and
\begin{equation}
\Delta^{2}_{NL}(k_{NL}) = f_{NL}(\Delta^{2}_{L}(k_L)),
\eql{PD}
\end{equation}
where the function $f_{NL}$ is given in Peacock and Dodds.

Armed with this transformation, 
we can invert it using the Newton-Raphson 
method to determine the linear power spectrum 
that would give rise to a given
non-linear one. This linear power spectrum can be evolved trivially from
early times until today, at each step of the way using \ec{PD} to
form the corresponding non-linear power spectrum. This gives a 
much more realistic $P(k,z)$, which can then be used to compute
the angular correlation function.

\begin{figure}
\centerline{\psfig{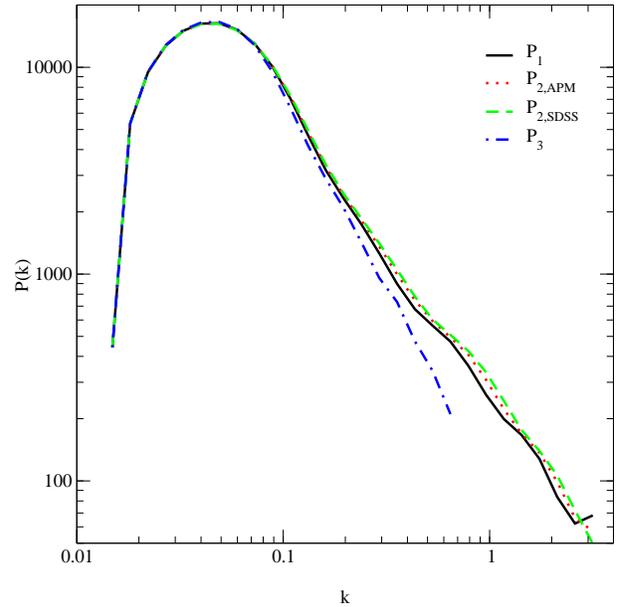}}
\centering
\caption{Present day values of three dimensional power spectra which might be 
inferred from angular data.  The solid line ($P_1$) is the present value of a 
spectrum which (i) evolves linearly and (ii) fits the angular data. The 
dashed and dotted curves ($P_2,APM$ and $P_2,SDSS$) are the present value of a 
spectrum which evolves non-linearly (i.e. realistically) and fits the agnular 
data. The dot-dashed curve $(P_3)$ is the linear spectrum which gives rise to 
the non-linear spectrum $P_2$. It should be used to fit cosmological 
parameters. } 
\label{fig:power}
\end{figure}

Figure~\ref{fig:power} shows three power spectra which might conceivably be
extracted from the APM survey. The three lines correspond
to three different power spectra, all shown at $z=0$:

\begin{itemize}

\item $P_1$ is the power spectrum one gets from the inversion
assuming linear scaling.

\item $P_2$ has the more realistic
scaling using the formalism of Peacock and Dodds, but
leads to a very similar $w(\theta)$ (see figure 2). We will discuss in the
next section how we arrived at this power spectrum. 

\item Finally $P_3$ is the linear
spectrum associated with the non-linear spectrum $P_2$. That is,
if the universe started with $P_3$
at early times (scaled back by $(1+z)^{-2}$), the non-linear power today
would be $P_2$.

\end{itemize}

\begin{figure}
%\vspace{.5 cm}
%\vspace{-2.5 cm}
\centerline{\psfig{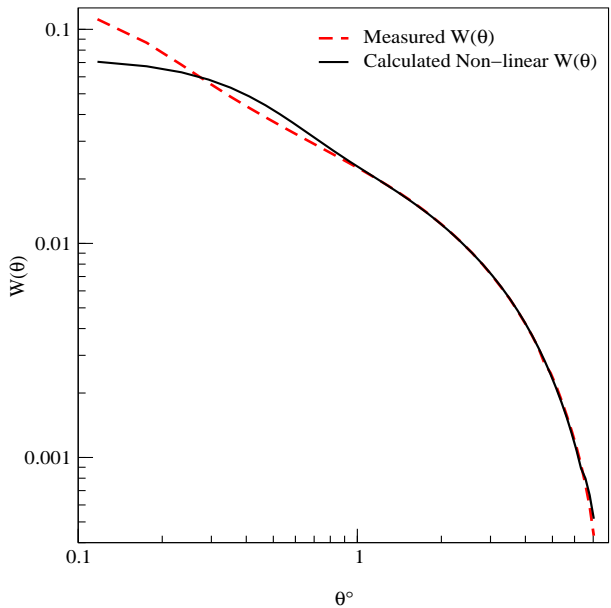}}
\centering
\caption{ Measured $w$ used to fit the three-dimensional power spectrum and $w$
calculated from the fit power spectrum using the non-linear method described 
herein.} 
\label{fig:wtheta}
%\centerline{\epsfxsize=8.truecm \epsfbox{PowerAPM.eps}}
%\centering
%\leavevmode\epsfxsize=8cm \epsfbox{Wtheta.eps}
%\caption[Fig1]{ \label{ Measured W used to fit the three-dimensional 
%power spectrum and W calculated from the fit power spectrum using 
%the non-linear method described herein.}  }
\end{figure}

The first of these, $P_1$, is what emerges from  a blind inversion
assuming $P\propto (1+z)^{-2}$. This is what is usually used to compare with 
theories. The second, $P_2$, is a much more accurate extraction of
the power spectrum today since we accounted for the non-linear evolution.
It clearly differs from $P_1$ at $k \ga 0.5$, so any attempt to use $P_1$
to constrain cosmological parameters will necessarily be inaccurate. The
only problem with $P_2$ is that we have not yet explained how we got it.
We'll do this in the next section. To do accurate parameter estimation, one
does a best fit to the data allowing for several free parameters and
evaluating the power spectrum at any point in parameter space with 
for example the BBKS (Bardeen et al. 1986) 
form or the output from CMBFAST (Seljak \& Zaldarriaga 1996). To do this comparison
properly, the power spectrum $P_3$ would need to be used, for the
codes compute the linear power spectra and $P_3$ is the linear spectra
corresponding to $P_2$. It has been common practice to neglect non-linear
effects and simply use $P_1$ to fit for cosmological parameters, neglecting
information on scales larger than some $k_{\rm max}$, typically chosen to be in the
range $0.1-0.2 h$ Mpc$^{-1}$. Since $P_1$
(the incorrect linear spectrum)
differs from $P_3$ (the correct linear spectrum) at wavenumbers $k$ even
smaller than $0.1 h$ Mpc$^{-1}$, it would be clearly be much better to find a
systematic way of obtaining $P_3$ or equivalently its non-linear counterpart,
$P_2$.

\section{The Weight Function}

We are almost ready to divulge the secret of how we got the spectrum $P_2$,
which we claim is a much better estimate of the power spectrum today than
is $P_1$. First, though, let's try to understand why the spectra $P_1$
and $P_2$ above differ. If the measurement was only of the power spectrum today,
then it wouldn't matter how the spectrum evolved with time beforehand: to fit the
angular correlation function, $P_1$ would have to be equal to $P_2$ today.
The difference between the two today, as reflected in figures 1 and 2, 
then must be due to the fact that the measurements are a weighted average
of the power spectrum over time. Different surveys will carry with
them different weight functions. Indeed, even different wavenumbers in the
same survey will have different weight functions. It is clearly very important
to understand and be able to compute the weight function.

\begin{figure}
%\vspace{5.5 cm}
%\vspace{-2.5 cm}
\centerline{\psfig{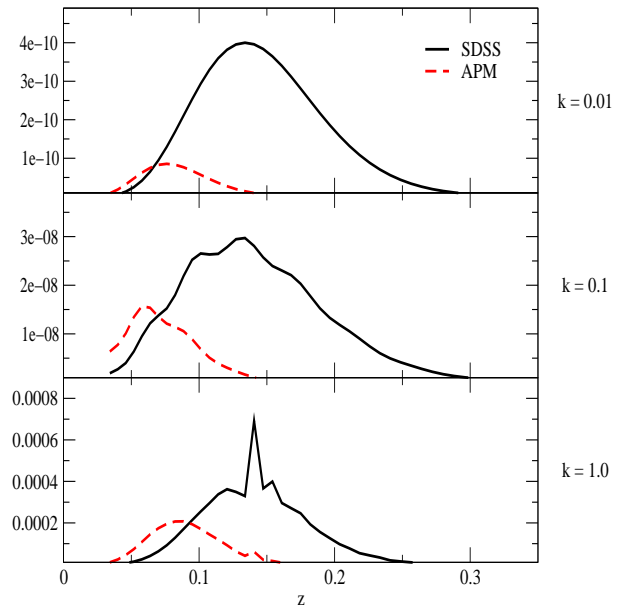}}
\centering
\caption{The weight function 
$W(k,z)$ for $k = 1.0, 0.1, 0.01 h$ Mpc$^{-1}$ for both the APM survey and 
SDSS, which goes wider and deeper.  This shows which values of $z$ contribute 
the most weight in the determination of the angular correlation function. The 
3D power spectrum inferred therefore is $P(k,z)$ at the peak of the wave 
function. }
\label{fig:weight}
%\centerline{\epsfxsize=8.truecm \epsfbox{PowerError.eps}}
%\centering
%\leavevmode\epsfxsize=8cm \epsfbox{PowerError.eps}
%\caption[Fig3]{ \label{ Power Error}  }
\end{figure}

The weight function can be computed by first forming
a $\chi^2$ from the observed $\hat w_i$'s and
the theoretical $w_i(P)$:
\begin{equation}
\chi^2 \equiv \left( \hat w_i - w_i(P) \right) C^{-1}_{ij} 
\left( \hat w_j - w_j(P) \right)
\end{equation}
where $C$ is the covariance matrix for $\hat w_i$.  If we want to
figure out how much weight the power spectrum at redshift
$z_\alpha$ and wavenumber $k_a$ 
contributes to the $\chi^2$, we need only compute the
second derivative of the $\chi^2$ with respect to $P_{a\alpha}$.
This gives the curvature, or the weight function:
\bea
W(k_a,z_\alpha) &\equiv & \frac{1}{2} {\partial^2\chi^2\over \partial P_{a\alpha}^2} \vs
&=& \sum_{i,j} {\cal K}_{ia\alpha} C^{-1}_{ij} 
{\cal K}_{ja\alpha}
.\eea
Plugging in from \ec{fullker}, we see that the weight function is
\bea
W(k,z) &=&  \left[ \frac{k \dk\dz}{2\pi}
\frac{x^{4}\psi^{2}  }{H F (1+z)^{6} } \right]^2 \\ \nonumber
& & \times \sum_i C^{-1}_{ij}  J_{0}(kx \sin \theta_i) J_{0}(kx \sin \theta_i).
\eql{weight}
\eea

The weight function is plotted in figure~\ref{fig:weight} for two surveys, APM
and the SDSS photometric survey\footnote{To do this, we have had to assume something
about the covariance matrix. We restricted ourselves to angular scales greater than
half a degree and assumed the covariance matrix was due solely to cosmic variance. We
computed this matrix assuming Gaussian statistics.}. For the former, which is shallower,
the weight function is peaked at $z\simeq 0.08$ and is fairly narrow.
%Note that larger wavenumbers have weight functions peaked at smaller
%redshifts. This makes sense intuitively: For a fixed range of angular
%scales, small physical scales correspond to
%objects closer to us, and hence lower redshifts. 
The weight function for SDSS
is also shown assuming galaxies can be extracted down to $22$nd magnitude.
As expected it peaks at higher redshift and is broader. For fixed $k$ we 
define $\zbar$ to be the redshift at which $W(k,z)$ peaks. 

The weight functions shown in figure~\ref{fig:weight} may
be somewhat suprising to those with knowledge of the
surveys. The median redshifts of these surveys are larger
than might be expected from consideration of
figure~\ref{fig:weight}. The discrepency can be attributed
to the fact that, for a given $k$, quite a bit of the
weight for the measurement of the power spectrum comes from
large angles (and therefore reshifts much smaller than the
median redshift). 

The weight function gives us a very clear way to think of the power
spectra extracted from the inversion. Recall that the inversion
techniques assume linear scaling. Since the weight function tells
us that a given $k-$~mode is mostly a measure of $P(k,\zbar)$, we should
scale back the inverted power spectrum to $\zbar$.  Then, if we want
the power spectrum today, we can scale forward with the non-linear
formulae. Indeed, we can now reveal that this is how we arrived at
$P_2$ in the previous section.

This suggests the following recipe for extracting a present day 3D power spectrum
from angular data:

\begin{itemize}

\item Assume linear scaling so that Eq.~(1) reduces to the much more managable
Eq.~(5). 

\item Invert to find $P_1$ today.

\item Scale $P_1(k,z=0)$ back linearly to $P_1(k,\zbar(k))
= P_1(k,z=0) (1+\zbar(k))^{-2}$. This scaled back spectrum
is a good estimator for the non-linear spectrum at $\zbar$.

\item Scale the spectrum obtained in the previous step non-linearly
to its present value, $P_2$. 

\item Find the underlying linear spectrum corresponding to $P_2$ and $P_3$, 
call it $P_3$. This can then be compared to linear models to extract
parameters. 

\end{itemize}

\section{Error on $P(k,\zbar)$}

There is one final loose end to tie up. The above prescription would be
exact if the weight function was a delta-function, infinitely sharp at
$\zbar$. Its finite width allows for the possibility that the scaling assumed
around $\zbar$ affects the measurement of $P(k,\zbar)$. There are several ways
we can test this possibility. The first is to look at the resultant angular
correlation functions from the two spectra. Figure 2 shows these and the
difference between the two for APM. It is encouraging that the difference is so
small. This suggests that the evolution of the power spectrum through
$\zbar$ is not very important; the measurements are simply of
$P(k,\zbar)$.

To test this further and to assign error bars due to the assumed scaling,
we can define
\be
\Delta_{\rm SC}^2(k) \equiv \sum_\alpha W(k,z_\alpha) (P_1(k,z_\alpha)
		- P_2(k,z_\alpha) )^2.
\eql{delta}\ee
Since $\sum_\alpha W(k,z_\alpha)$ is the weight of the measurement,
it is the inverse of the square of the error on the measurement of
$P(k)$. Therefore, $\Delta_{\rm SC}$ can be thought of as the ratio of
the error due to the linear scaling assumption used in the inversion
process to the overall error in the power spectrum. If $\Delta_{\rm SC}$ is 
much less than one, then we need not worry about the scaling assumption. 
Figure~\ref{fig:error} shows that, for surveys with the depth of APM or even 
SDSS, $\Delta_{\rm SC}$ is indeed quite small for $k \la 1 h$ Mpc$^{-1}$.
The broader weight function of SDSS leads to a larger scaling error.
One might argue that, even for $k$ as low as $0.3 h$ Mpc$^{-1}$, the statistical error
on the power spectrum is an underestimate. Taking $\Delta^2_{\rm SC}$ to be $0.2$
there leads suggests (assuming errors add in quadrature) 
that the linear evolution assumption increases the
errors by a factor of $\sqrt{1 + 0.2}$, about ten percent. And of course, at higher
wavenumbers the error bars get even larger.

\begin{figure}
%\vspace{5.5 cm}
%\vspace{-2.5 cm}
\centerline{\psfig{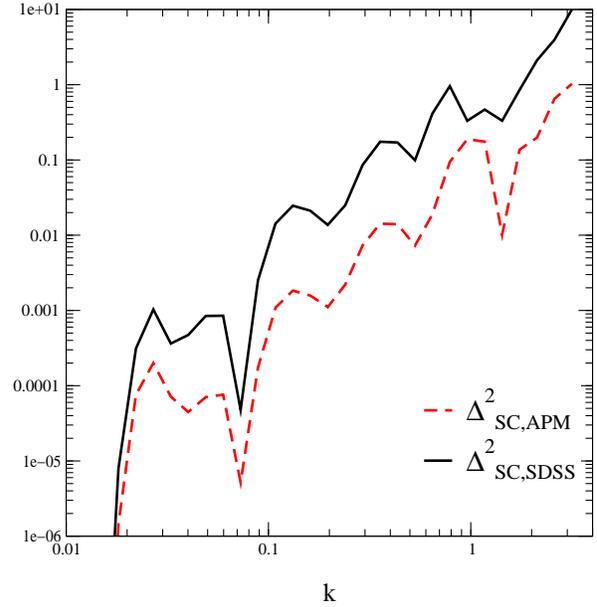}}
\centering
\caption{ The scaling error $\Delta^{2}_{\rm SC}$ for APM and SDSS. $\Delta_{\rm SC}$
is a measure of the error induced by assuming linear scaling to the statistical error.}
\label{fig:error}
%\centerline{\epsfxsize=8.truecm \epsfbox{PowerError.eps}}
%\centering
%\leavevmode\epsfxsize=8cm \epsfbox{PowerError.eps}
%\caption[Fig3]{ \label{ Power Error}  }
\end{figure}

\section{Conclusions}

The inversion of the angular correlation function gives a measure of the
three-dimensional power spectrum at redshift $\zbar$, where $\zbar$
is the place where $W(k,z)$ in \ec{weight} peaks. A simple way to obtain
an estimate for $P(k,\zbar)$ is to assume linear scaling of the power
spectrum, invert the kernel, and scale back the power spectrum to
$\zbar$. An estimate in the error incurred by this procedure is given by
\ec{delta}. For current, and even future surveys, the error is small
for $k < 0.5 h$ Mpc$^{-1}$ (but might be much more significant for deep surveys). 
On larger scales, the error is larger.
This is not necessarily a bad thing: it is an indication that the survey is
sensitive to the evolution of the power spectrum. In fact, the tools developed here
could be applied to help plan surveys or devise optimal strategies for breaking a
survey into subsets. 
One could compute the weight function for a given subset of data (e.g. a given
magnitude slice) and choose a
different subset whose weight function peaks far away in redshift. The width of the
weight function in SDSS suggests that this may already be possible.

We have not dealt at all with the possibility of using photometric
redshifts to learn more about the evolution of the power spectrum.
And we have completely ignored the issue of how the galaxies are
biased with respect to the matter. There has been much activity in
both of these fields over the past several years, which should help 
extract even more useful information from angular surveys.

{\bf Acknowledgments}
This work is supported by NASA Grant NAG 5-7092 and the DOE.

\section{References}
\def\refe {\par \hangindent=.7cm \hangafter=1 \noindent}
\def\apj { ApJ }
\def\astroph{{\tt astro-ph/}} 
\def\aap {A \& A }
\def\ajs{ ApJS }
\def\aj{AJ}
\def\prd{Phys ReV D}
\def\apjs{ ApJS }
\def\mnras { MNRAS }
\def\apjl { Ap. J. Let. }

\refe Baugh, C.M., Efstathiou, G., 1993, \mnras 265, 145 
\refe Baugh, C.M., Efstathiou, G., 1994, \mnras 267, 323
\refe Bardeen, J.~M., Bond, J.~R., Kaiser, N., \& Szalay, A.~S.,
1986, \apj, 304, 15
\refe Dodelson, S. \& Gazta\~{n}aga, E., 1999, \astroph 9906289
\refe Eisenstein, D.~J. \& Zaldarriaga, M., 1999, \astroph 9912149
\refe Gazta\~{n}aga, E. \& Baugh, C.M., 1998, \mnras, 294, 229 
\refe  Maddox, S.~J., Efstathiou, G., Sutherland, W.~J., \& Loveday, L. 1990, \mnras, 242, 43P
\refe Peacock, J. A. \& Dodds, S. J., 1996, \mnras, 280, 19
\refe Seljak, U. \& Zaldarriaga, M., 1996, \apj 469, 437

\end{document}